# Non-dipole angular anisotropy parameters of photoelectrons from semi-filled shell atoms

## M. Ya. Amusia[1, 2] and L. V. Chernysheva[2]

[1]Racah Institute of Physics, The Hebrew University, Jerusalem 91904, Israel
[2]Ioffe Physical-Technical Institute, St.-Petersburg 194021, Russia

**Abstract.** We present the results of calculations of outer and next to the outer shell non-dipole angular anisotropy parameters of photoelectrons for semi-filled shell atoms in the Hartree-Fock (HF) one-electron approximation and with account of inter-electron correlations in the frame of the Spin Polarized Random Phase Approximation with Exchange (SP RPAE). We demonstrate for the first time that this characteristic of photoionization process is essentially sensitive to the fact whether the photoelectron has the same or opposite spin orientation to that of the semi-filled shell.

## 1. Introductory remarks

The non-dipole parameters of the photoelectron angular distribution are relatively small quantities, in general smaller than the dipole parameter by the factor $v_i / c << 1$, where $v_i$ is the mean speed of the ionized electron and $c$ is the speed of light. However, as a result of construction of high intensity continuous spectrum radiation devices, namely synchrotrons and storage rings, the non-dipole parameters became measurable quantities.

These parameters are of great interest including information not only on dipole transition matrix elements but also on quadrupole ones. Therefore they can demonstrate the existence of invisible in the absolute cross-section quadrupole resonances and shed light on the behavior of photoelectron scattering phases that corresponds not only to $l \to l \pm 1$ but also to $l \to l, l \pm 2$ transitions, where $l$ is the angular momentum of the ionized electron.

The expression for differential in energy cross-section is well known since long ago [1, 2, 3], as well as the concrete expressions for the dipole parameter $\beta_{nl}(\omega)$ [3, 4] and non-dipole parameters $\gamma_{nl}(\omega)$ and $\eta_{nl}(\omega)$ [4, 5] for photoionization of an $nl$ subshell (where $\omega^{*/}$ is the photon frequency) in the frame of one-electron Hartree-Fock (HF) approximation and Random Phase Approximation with Exchange (RPAE), which takes into account the most important part of the inter-electron correlations.

Extensive calculations of non-dipole parameters were performed mainly for closed shell ions and atoms [6, 7]. Special attention was given to noble gases [8] and negative ions [9]. There exist several papers dedicated to the results of non-dipole parameters measurements (see e.g. [10-12]). However, almost nothing was done in this direction for open shell atoms, particularly with accounting for the role of electron correlations. The latter are especially important in these atoms.

---

*/ Note, that we use the atomic system of units, $e = m_e = \hbar = 1$ in this paper.

The general consideration of open shell atoms in the frame of RPAE is rather complex. But it was demonstrated quite a while ago that semi-filled shell atom, being in one sense open shell atoms can be treated in the RPAE frame as closed shell atoms with two types, namely "*up*" and "*down*" electrons, thus called according to their spin direction. The corresponding generalization of RPAE can be found in [13].

## 2. Main formulas

In this section we will present the main formulas used in this paper. Since they are known, we will limit ourselves to the final results only, omitting the corresponding derivations. The following relation gives the differential in angle photoionization cross-section by non-polarized light [4]:

$$\frac{d\sigma_{nl}(\omega)}{d\Omega} = \frac{\sigma_{nl}(\omega)}{4\pi}[1 - \frac{\beta_{nl}(\omega)}{2}P_2(\cos\theta) + \kappa\gamma_{nl}(\omega)P_1(\cos\theta) + \kappa\eta_{nl}(\omega)P_3(\cos\theta)], \qquad (1)$$

where $\kappa = \omega/c$, $P_{1,2,3}(\cos\theta)$ are the Legendre polynomials, $\theta$ is the angle between photon and photoelectron momenta, $\beta_{nl}(\omega)$ is the dipole, while $\gamma_{nl}(\omega)$ and $\eta_{nl}(\omega)$ are so-called non-dipole angular anisotropy parameters.

In experiment, usually sources of linearly polarized radiation are used. In this case instead of (1) another form of angular distribution is more convenient [14,15]:

$$\frac{d\sigma_{nl}(\omega)}{d\Omega} = \frac{\sigma_{nl}(\omega)}{4\pi}\{1 + \beta_{nl}(\omega)P_2(\cos\vartheta) + [\delta_{nl}^C(\omega) + \gamma_{nl}^C(\omega)\cos^2\vartheta]\sin\vartheta\cos\Phi\}. \qquad (2)$$

Here $\vartheta$ is the polar angle between the vectors of photoelectron's velocity $\vec{v}$ and photon's polarization $\vec{e}$, while $\Phi$ is the azimuth angle determined by the projection of $\vec{v}$ in the plane orthogonal to $\vec{e}$ that includes the vector of photon's velocity. The non-dipole parameters in (1) and (2) are connected by simple relations [16]

$$\gamma_{nl}^C/5 + \delta_{nl}^C = \kappa\gamma_{nl}, \qquad\qquad \gamma_{nl}^C/5 = -\kappa\eta_{nl}. \qquad (3)$$

The presented below results of calculations of non-dipole parameters are obtained using both expression, namely (1) and (2).

There are two possible dipole transitions from subshell $l$, namely $l \to l \pm 1$ and three quadrupole transitions $l \to l \pm 2, l$. Corresponding general expressions for $\beta_{nl}(\omega)$, $\gamma_{n\ell}(\omega)$ and $\eta_{n\ell}(\omega)$ are rather complex and expressed via the matrix elements of these transitions. In one –electron Hartree-Fock approximation they can be presented as [3, 8]:

$$\beta_{nl}(\omega) = \frac{1}{(2l+1)[(l+1)d_{l+1}{}^2 + ld_{l-1}{}^2]}[(l+2)(l+1)d_{l+1}{}^2 + l(l-1)d_{l-1}{}^2 - 6(l+1)ld_{l+1}d_{l-1} \times$$

$$\cos(\delta_{l+1} - \delta_{l-1})].$$

$$(4)$$

$$\gamma_{nl}(\omega) = \frac{3}{5[ld_{l-1}^2 + (l+1)d_{l+1}^2]} \left\{ \frac{l+1}{2l+3}[3(l+2)q_{l+2}d_{l+1}\cos(\delta_{l+2} - \delta_{l+1}) - lq_l d_{l+1} \times \right.$$

$$\left. \cos(\delta_{l+2} - \delta_{l+1})] - \frac{l}{2l+1}[3(l-1)q_{l-2}d_{l-1}\cos(\delta_{l-2} - \delta_{l-1}) - (l+1)q_l d_{l-1}\cos(\delta_l - \delta_{l-1})] \right\}, \quad (5)$$

$$\eta_{nl}(\omega) = \frac{3}{5[ld_{l-1}^2 + (l+1)d_{l+1}^2]} \left\{ \frac{(l+1)(l+2)}{(2l+1)(2l+3)}q_{l+2}[5ld_{l-1}\cos(\delta_{l+2} - \delta_{l-1}) - \right.$$

$$- (l+3)d_{l+1}\cos(\delta_{l+2} - \delta_{l-1})] - \frac{(l-1)l}{(2l+1)(2l+1)}q_{l-2} \times$$

$$\times [5(l+1)d_{l+1}\cos(\delta_{l-2} - \delta_{l+1}) - (l-2)d_{l-1}\cos(\delta_{l-2} - \delta_{l-1})] +$$

$$+ 2\frac{l(l+1)}{(2l-1)(2l+3)}q_l[(l+2)d_{l+1}\cos(\delta_l - \delta_{l+1}) - (l-1)d_{l-1}\cos(\delta_l - \delta_{l-1})] \right\}. \quad (6)$$

Here $\delta_{l'}$ are photoelectrons' scattering phases; the following relation gives the matrix elements $d_{l\pm1}$ in the so-called *r*-form

$$d_{l\pm1} \equiv \int_0^\infty P_{nl}(r) r P_{\varepsilon l\pm1}(r) dr, \quad (7)$$

where $P_{nl}(r)$, $P_{\varepsilon l\pm1}(r)$ are the radial Hartree-Fock one-electron wave functions of *nl* discrete level and $\varepsilon l'$ - in continuous spectrum, respectively. The following relation gives the quadrupole matrix elements

$$q_{l\pm2,0} \equiv \frac{1}{2}\int_0^\infty P_{nl}(r) r^2 P_{\varepsilon l\pm2,l}(r) dr. \quad (8)$$

In order to take into account the Random Phase Approximation with Exchange (RPAE) [3, 17] multi-electron correlations, one has to perform the following substitutions in the expressions for $\beta_{nl}(\omega)$, $\gamma_{nl}(\omega)$ and $\eta_{nl}(\omega)$ [8]:

$$d_{l+1}d_{l-1}\cos(\delta_{l+1} - \delta_{l-1}) \rightarrow [(\operatorname{Re} D_{l+1}\operatorname{Re} D_{l-1} + \operatorname{Im} D_{l+1}\operatorname{Im} D_{l-1})\cos(\delta_{l+1} - \delta_{l-1}) -$$

$$- (\operatorname{Re} D_{l+1}\operatorname{Im} D_{l-1} - \operatorname{Im} D_{l+1}\operatorname{Re} D_{l-1})\sin(\delta_{l+1} - \delta_{l-1})] \quad (9)$$

$$d_{l\pm1}q_{l\pm2,0}\cos(\delta_{l\pm2,0} - \delta_{l\pm1}) \rightarrow [(\operatorname{Re} D_{l\pm1}\operatorname{Re} Q_{l\pm2,0} + \operatorname{Im} D_{l\pm1}\operatorname{Im} Q_{l\pm2,0})\cos(\delta_{l\pm2,l} - \delta_{l\pm1}) -$$

$$- (\operatorname{Re} D_{l\pm1}\operatorname{Im} Q_{l\pm2,0} - \operatorname{Im} D_{l\pm1}\operatorname{Re} Q_{l\pm2,0})\sin(\delta_{l\pm2,l} - \delta_{l\pm1})], \quad (10)$$

$$d_{l\pm1}^2 \rightarrow \operatorname{Re} D_{l\pm1}^2 + \operatorname{Im} D_{l\pm1}^2. $$

The following is the RPAE equation for the dipole matrix elements

$$\langle v_2|D(\omega)|v_1\rangle = \langle v_2|d|v_1\rangle + \sum_{v_3,v_4} \frac{\langle v_3|D(\omega)|v_4\rangle (n_{v_4} - n_{v_3})\langle v_4 v_2|U|v_3 v_1\rangle}{\varepsilon_{v_4} - \varepsilon_{v_3} + \omega + i\eta(1 - 2n_{v_3})}, \quad (11)$$

where

$$\langle \nu_1\nu_2|\hat{U}|\nu_1'\nu_2'\rangle \equiv \langle \nu_1\nu_2|\hat{V}|\nu_1'\nu_2'\rangle - \langle \nu_1\nu_2|\hat{V}|\nu_2'\nu_1'\rangle. \tag{12}$$

Here $\hat{V} \equiv 1/|\vec{r} - \vec{r}'|$ and $\nu_i$ is the total set of quantum numbers that characterize a HF one-electron state on discrete (continuum) levels. That includes principal quantum number (energy), angular momentum, its projection and the projection of the spin. The function $n_{\nu_i}$ (the so-called step-function) is equal to 1 for occupied and 0 for vacant states.

The dipole matrix elements $D_{l\pm1}$ are obtained by solving the radial part of the RPAE equation (11). As to the quadrupole matrix elements $Q_{l\pm2,0}$, they are obtained by solving the radial part of the RPAE equation, similar to (11)

$$\langle \nu_2|Q(\omega)|\nu_1\rangle = \langle \nu_2|\hat{q}|\nu_1\rangle + \sum_{\nu_3,\nu_4} \frac{\langle \nu_3|Q(\omega)|\nu_4\rangle(n_{\nu_4}-n_{\nu_3})\langle \nu_4\nu_2|U|\nu_3\nu_1\rangle}{\varepsilon_{\nu_4}-\varepsilon_{\nu_3}+\omega+i\eta(1-2n_{\nu_3})}. \tag{13}$$

Here in $r$-form one has $\hat{q} = r^2 P_2(\cos\theta)$. Equations (11, 13) are solved numerically using the procedure discussed at length in [17].

The expressions (4) - (6) are essentially simplified for $s$-subshells that are the main subject of this paper. They become equal to

$$\beta_{n0}(\omega) = 2, \tag{14}$$

$$\gamma_{n0}(\omega) = -\eta_{n0}(\omega) = \frac{6}{5d_1^2} d_1 q_2 \cos(\delta_d - \delta_p), \tag{15}$$

$$\gamma_{n0}^C(\omega) = \frac{6\kappa}{d_1^2} d_1 q_2 \cos(\delta_d - \delta_p), \tag{16}$$

where $d_1(q_2)$ are the dipole (quadrupole) matrix elements $\langle \varepsilon, p(d)|d(q)|ns\rangle$, and $\delta_{p(d)}$ - are the scattering phases of p(d)-photoelectron. In RPAE the expression for $\gamma_{n0}(\omega) = -\eta_{n0}(\omega)$ can be obtained from (10) by substituting

$$d_1 q_2 \cos(\delta_2 - \delta_1) \rightarrow [(\operatorname{Re}D_1 \operatorname{Re}Q_2 + \operatorname{Im}D_1 \operatorname{Im}Q_2)\cos(\delta_2 - \delta_1) -$$
$$- (\operatorname{Re}D_1 \operatorname{Im}Q_2 - \operatorname{Im}D_1 \operatorname{Re}Q_2)\sin(\delta_2 - \delta_1)],$$
$$d_1^2 \rightarrow \operatorname{Re}D_1^2 + \operatorname{Im}D_1^2. \tag{17}$$

In this paper we investigate semi-filled shell atoms. As it was demonstrated in [13], they can be treated as objects that consists of two types of different particles, namely "*up*" and "*down*" electrons, according to their spin projection, $\uparrow$ and $\downarrow$, with no exchange between them. In such case the ordinary HF approximation has to be replaced by Spin-Polarized Hartree-Fock (SP HF) and RPAE has to be replaced by SP RPAE. Most conveniently this can be done in the operator, symbolic form.

The dipole RPAE equation (11) in the operator form looks like [18]

$$\hat{D}(\omega) = d + \hat{D}(\omega)\hat{\chi}(\omega)\hat{U}, \tag{18}$$

where so-called electron-vacancy propagator $\hat{\chi}(\omega)$ is given by the following relation

$$\hat{\chi}(\omega) = \hat{1}/(\omega - \hat{H}_{ev}) - \hat{1}/(\omega + \hat{H}_{ev}),  \tag{19}$$

where $\hat{H}_{ev}$ is the electron – vacancy Hartree – Fock Hamiltonian.

As a result of the difference of "*up*" and "*down*" states, all the subshells become separated into two levels. For semi-filled atoms all these levels are totally filled with "*up*", $\uparrow$ and "*down*", $\downarrow$ electrons [13]. The corresponding generalization of RPAE is easy to achieve using instead of (18) the following equation that has a matrix form:

$$\left(\hat{D}_{\uparrow}(\omega)\hat{D}_{\downarrow}(\omega)\right) = \left(\hat{d}_{\uparrow}(\omega)\hat{d}_{\downarrow}(\omega)\right) + \left(\hat{D}_{\uparrow}(\omega)\hat{D}_{\downarrow}(\omega)\right) \times \begin{pmatrix} \hat{\chi}_{\uparrow\uparrow} & 0 \\ 0 & \hat{\chi}_{\downarrow\downarrow} \end{pmatrix} \times \begin{pmatrix} \hat{U}_{\uparrow\uparrow} \hat{V}_{\uparrow\downarrow} \\ \hat{V}_{\downarrow\uparrow} \hat{U}_{\downarrow\downarrow} \end{pmatrix}  \tag{20}$$

Here the arrows $\uparrow (\downarrow)$ denote the one-electron functions related to "*up*"("*down*") electrons, respectively. Note the presence of pure direct (without exchange) interaction $\hat{V}$ between electron-vacancy pairs created by exciting either "*up*" or "*down*" electrons.

The generalization of (13) is similar

$$\left(\hat{Q}_{\uparrow}(\omega)\hat{Q}_{\downarrow}(\omega)\right) = \left(\hat{q}_{\uparrow}(\omega)\hat{q}_{\downarrow}(\omega)\right) + \left(\hat{Q}_{\uparrow}(\omega)\hat{Q}_{\downarrow}(\omega)\right) \times \begin{pmatrix} \hat{\chi}_{\uparrow\uparrow} & 0 \\ 0 & \hat{\chi}_{\downarrow\downarrow} \end{pmatrix} \times \begin{pmatrix} \hat{U}_{\uparrow\uparrow} \hat{V}_{\uparrow\downarrow} \\ \hat{V}_{\downarrow\uparrow} \hat{U}_{\downarrow\downarrow} \end{pmatrix}  \tag{21}$$

Since "*up*" or "*down*" electrons are different, the angular distribution for them is also different, thus acquire along with $nl$ an additional lower index $\uparrow(\downarrow)$. The same is correct for the absolute cross-section that becomes $\sigma_{nl\uparrow(\downarrow)}(\omega)$ and angular anisotropy parameters that become $\beta_{nl\uparrow(\downarrow)}(\omega)$, $\gamma_{nl\uparrow(\downarrow)}(\omega)$ and $\eta_{nl\uparrow(\downarrow)}(\omega)$. All they are expressed via spin-polarized matrix elements $d_{l'\uparrow(\downarrow)}$, $q_{l'\uparrow(\downarrow)}$ and phases $\delta_{l'\uparrow(\downarrow)}$. The matrix elements $d_{l'\uparrow(\downarrow)}$, $q_{l'\uparrow(\downarrow)}$ are determined by relations, similar to (7) and (8) that correspond to "*up*" or "*down*" wave functions

$$d_{l\pm 1\uparrow(\downarrow)} \equiv \int_0^\infty P_{nl\uparrow(\downarrow)}(r) r P_{\varepsilon l\pm 1\uparrow(\downarrow)}(r) dr, \qquad q_{l\pm 2,0\uparrow(\downarrow)} \equiv \frac{1}{2}\int_0^\infty P_{nl\uparrow(\downarrow)}(r) r^2 P_{\varepsilon l\pm 2,0\uparrow(\downarrow)}(r) dr.  \tag{22}$$

To obtain $\beta_{nl\uparrow(\downarrow)}(\omega)$, $\gamma_{nl\uparrow(\downarrow)}(\omega)$ and $\eta_{nl\uparrow(\downarrow)}(\omega)$ in SP RPAE, substitutions similar to (10) has to be done, using solutions (20) and (21).

### 3. Results of calculations

We present here results of calculations of non-dipole photoelectron angular anisotropy parameters for outer s-electrons along with the nearest semi-filled subshells of Li ($^2S_{1/2}$), P ($^4S_{3/2}$), Cr ($^7S_3$), Cr$^*$ ($^5S_2$), Mn ($^6S_{5/2}$), and Eu ($^8S_{7/2}$).

The level structure of these atoms looks as follows:

Li: $\underline{\underline{1s \uparrow, \downarrow}}; \underline{\underline{2s \uparrow}}$,

P: $1s \uparrow, \downarrow; 2s \uparrow, \downarrow; 2p^3 \uparrow, \downarrow; \underline{3s \uparrow, \downarrow}; \underline{\underline{3p^3 \uparrow}}$,

Cr: $1s \uparrow, \downarrow; 2s \uparrow, \downarrow; 2p^3 \uparrow, \downarrow; 3s \uparrow, \downarrow; 3p^3 \uparrow, \downarrow; \underline{3d^5 \uparrow}; \underline{\underline{4s \uparrow}}$,

Cr$^*$: $1s \uparrow, \downarrow; 2s \uparrow, \downarrow; 2p^3 \uparrow, \downarrow; 3s \uparrow, \downarrow; 3p^3 \uparrow, \downarrow; \underline{3d^5 \uparrow}; \underline{\underline{4s \downarrow}}$

Mn: $1s \uparrow, \downarrow; 2s \uparrow, \downarrow; 2p^3 \uparrow, \downarrow; 3s \uparrow, \downarrow; 3p^3 \uparrow, \downarrow; \underline{3d^5 \uparrow}; \underline{\underline{4s \uparrow, \downarrow}}$,

Eu :

$1s \uparrow, \downarrow; 2s \uparrow, \downarrow; 2p^3 \uparrow, \downarrow; 3s \uparrow, \downarrow; 3p^3 \uparrow, \downarrow; 3d^5 \uparrow; 4s \uparrow, \downarrow; 4p^3 \uparrow, \downarrow; 4d^5 \uparrow, \downarrow; 5s \uparrow, \downarrow 5p^3 \uparrow, \downarrow;$
$\underline{4f^7 \uparrow}; \underline{\underline{6s \uparrow, \downarrow}}$                                                            .

With a single line we denote the semi-filled subshell that acts upon outer s-electrons that are emphasized by a double line.

The angular anisotropy parameters $\gamma^C(\omega)$ in HF approximation were obtained using (16), while the RPAE corrections were included using substitution (17). For all but Li atom the non-dipole parameters are dependent upon the term of the ion in the final state. This term corresponds to a spin that is smaller by 1/2 than the spin of the initial state, if an "up" electron is eliminated, and bigger by 1/2, if a "down" electron is eliminated.

In these notations "down" means final state term for s-electrons photoionization for P - $^5S_2$, Cr$^*$ - $^6S_{5/2}$, Mn - $^7S_3$, Eu - $^9S_4$, while "up" means for P - $^3S_1$, Cr - $^6S_{5/2}$, Mn - $^5S_2$, Eu - $^7S_3$. Note that for Cr and Cr$^*$ the final state term after outer s-electron photoionization remains, naturally, the same.

The Figure 1 for Li $\gamma^C_{1s}$ shows no influence of the final state term. On the contrary, the onset of 1s threshold affects $\gamma^C_{2s}$ , presented in Figure 2, leading to a prominent maximum at about 50 eV. The situation in P, as is seen from Figure 3, is completely different: $\gamma^C_{3s\uparrow}$ and $\gamma^C_{3s\downarrow}$ have not too much in common, particularly close to respective rather different thresholds. In both of them the effect of 3p-electrons are quite impressive. As to depicted in Figure 4 $\gamma^C_{3p\uparrow}$, it has a strong maximum at 31 eV, while $\delta^C_{3p\uparrow}$ being by a factor of ten smaller than $\gamma^C_{3p\uparrow}$, has at 32 eV a deep minimum. Note that $\gamma^C_{3p\uparrow}$ and $\delta^C_{3p\uparrow}$ are almost not affected by electron correlations.

As it's seen from Figure 5 for Cr the difference between $\gamma^C_{4s\uparrow}$ and $\gamma^C_{4s\downarrow}$ is big and impressive, particularly close to thresholds that are almost the same for "up" and "down" cases. The role of electron correlations in $\gamma^C_{3d\uparrow}$ is negligible according to Figure 6. As it is seen from respective Figures 7 and 8, the situation is almost the same in Mn.

In Eu, as it's presented in Figure 9, the difference between $\gamma^C_{6s\downarrow}$ and $\gamma^C_{6s\uparrow}$ is very big within 10 eV region near threshold. At higher energies they are close to each other and very small.

Entirely, we detect a strong dependence of the non-dipole angular anisotropy parameters for outer s-electrons in semi-filled shell atoms upon the tern of the residual ion. This effect can be detected experimentally since the term-dependence shows up in rather prominent corrections that are within the already achieved experimental accuracy (see e.g. [10]). Note, that in spite of the fact that the non-dipole anisotropy parameter is usually for low enough photon energies much smaller than the dipole one (equal to 2 for s-subshells), it is quite measurable experimentally almost from the threshold even for He [19].

Till now the main attention in the studies of non-dipole parameters were concentrated mainly on noble gas atoms, particularly on their inner and intermediate subshells. We do hope that the results presented in this paper will stimulate the efforts of experimentalists that will pay more attention to investigation of the non-dipole parameters of the outer subshells of not only noble gas atoms.

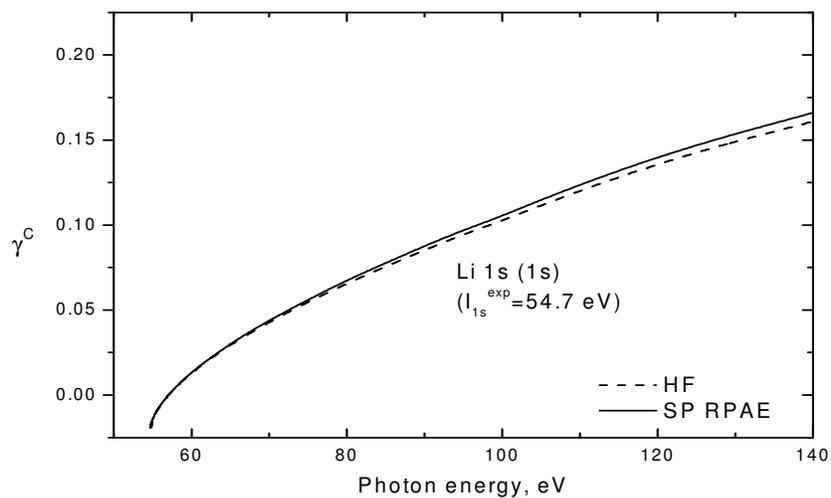

**Figure 1**. Non-dipole parameter $\gamma^c$ for 1s-electrons in Li.
Solid line is SP RPAE and dashed line is HF results, respectively.

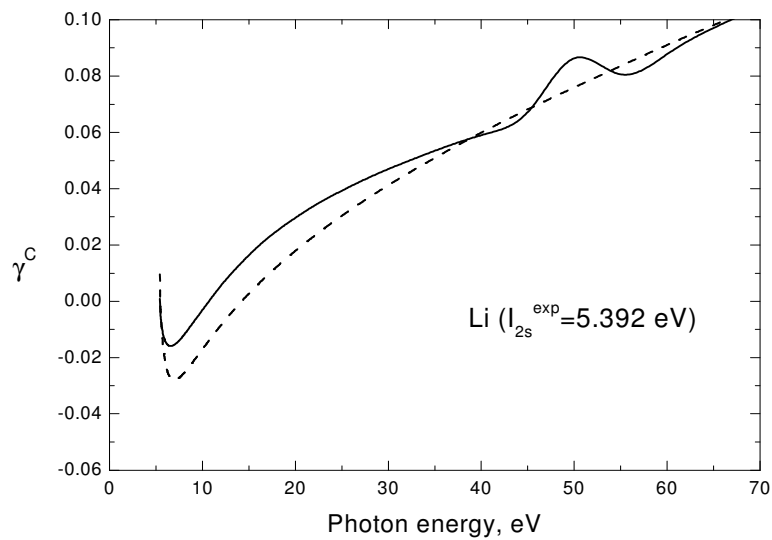

**Figure 2**. Non-dipole parameter $\gamma^c$ for 2s-electrons in Li.
Solid line is SP RPAE and dashed line is HF results, respectively.

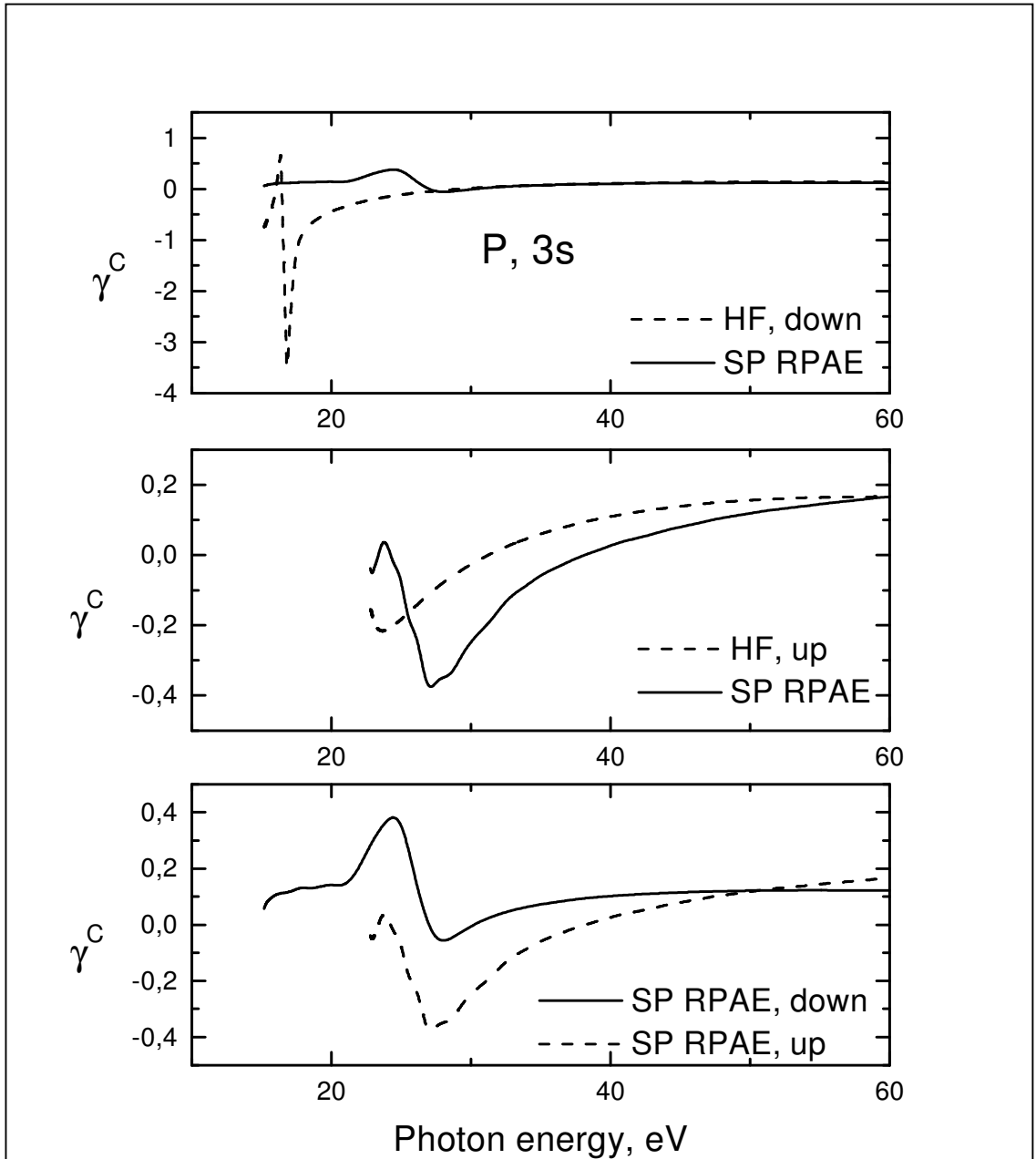

**Figure 3**. Non-dipole parameter $\gamma^c$ for 3s-electrons in P.

Here 'down' means $^5S_2$, 'up' - $^3S_1$, final state term of the ion. SP RPAE takes into account the interaction of 3s with 3p-electrons. Solid line is SP RPAE, dashed line is HF results, respectively. $I_{3s\uparrow}$= 1.67 Ry, $I_{3s\downarrow}$=1.11 Ry.

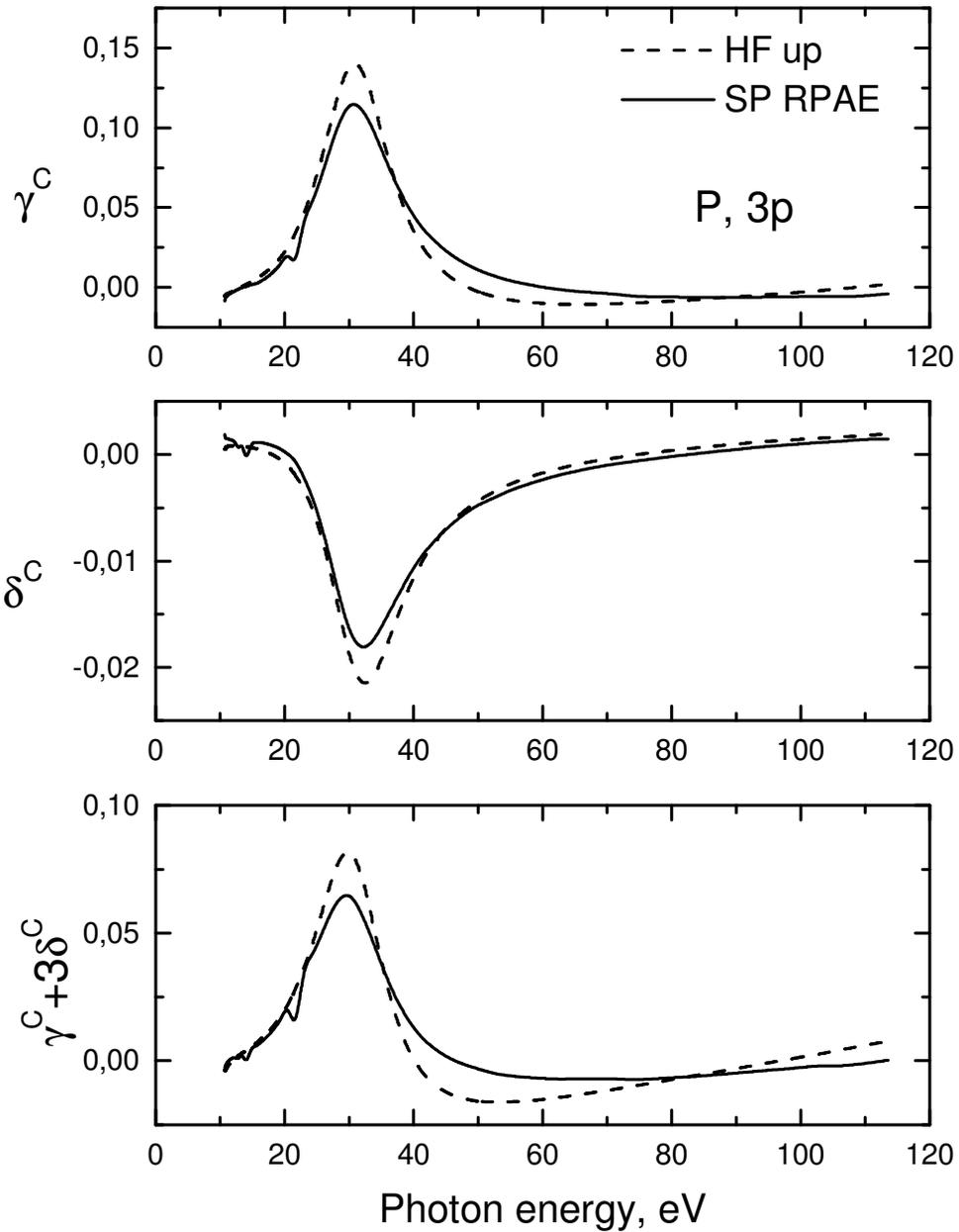

**Figure 4.** Non-dipole parameters for $3p\uparrow$-electrons in P.

Solid line is SP RPAE and dashed line is HF results, respectively $I_{3p\uparrow}=0.784$ Ry.

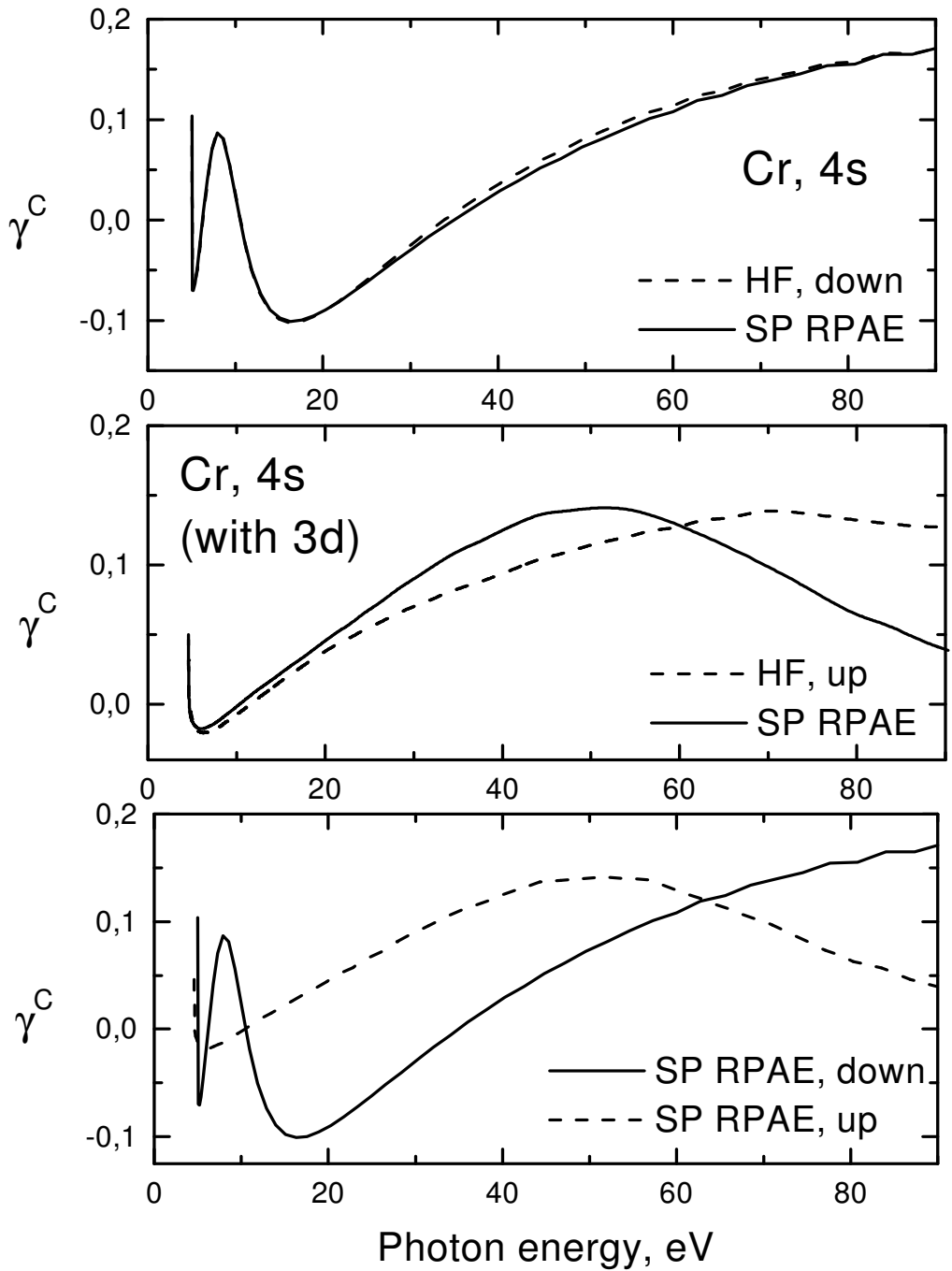

**Figure 5.** Non-dipole parameter $\gamma^c$ for 4s-electron in the excited and ground state of Cr with terms $^5S_2$ and $^7S_3$, respectively. SP RPAE takes into account the interaction of 4s with 3d-electrons. Solid line is SP RPAE and dashed line is HF results, respectively. $I_{4s\uparrow} = 0.335$ Ry, $I_{4s\downarrow} = 03687$ Ry.

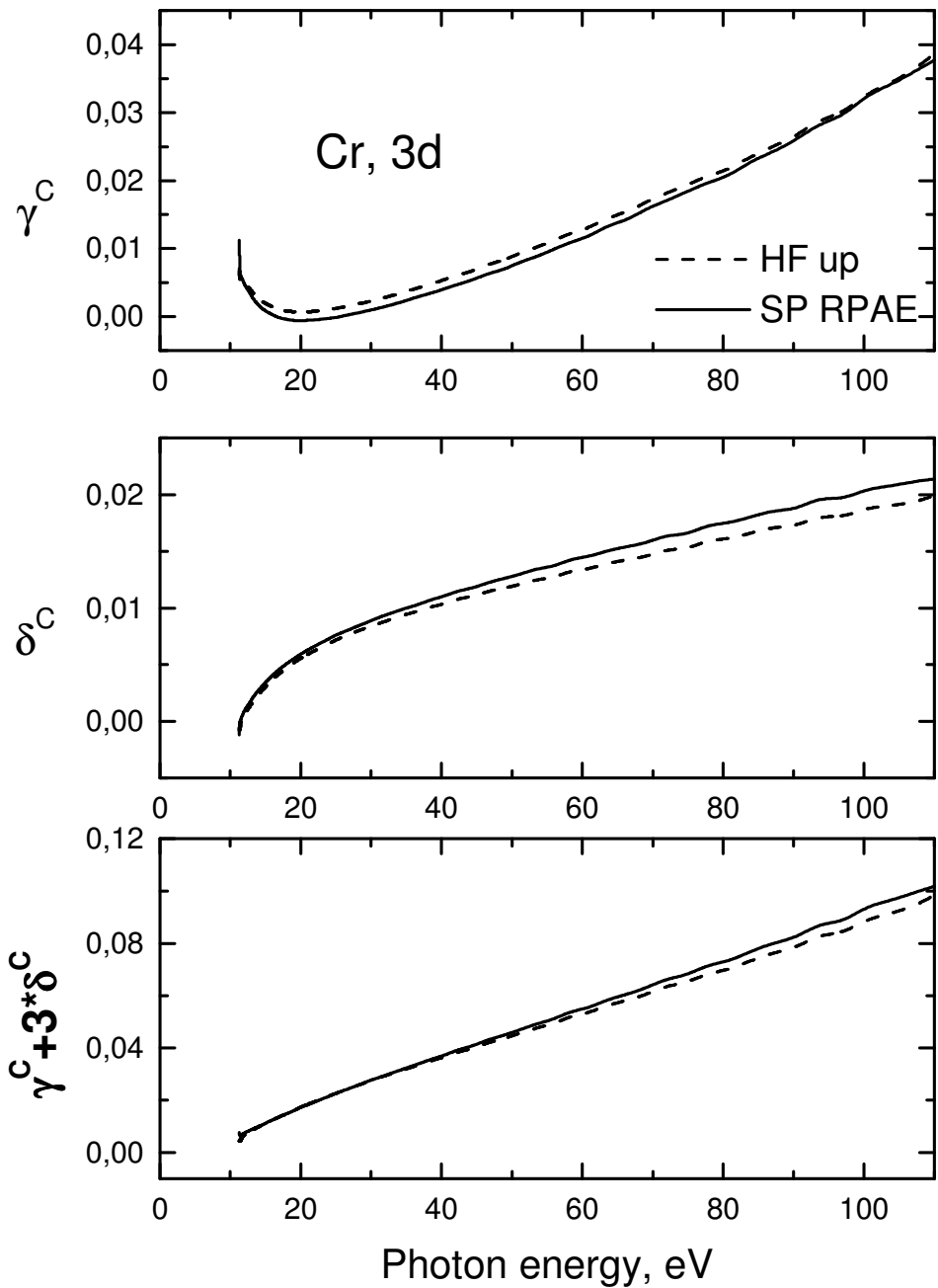

**Figure 6.** Non-dipole parameters for $3d \uparrow$-electrons in Cr.
Solid line is SP RPAE and dashed line is HF results, respectively.
$I_{3d\uparrow} = 0.831$ Ry.

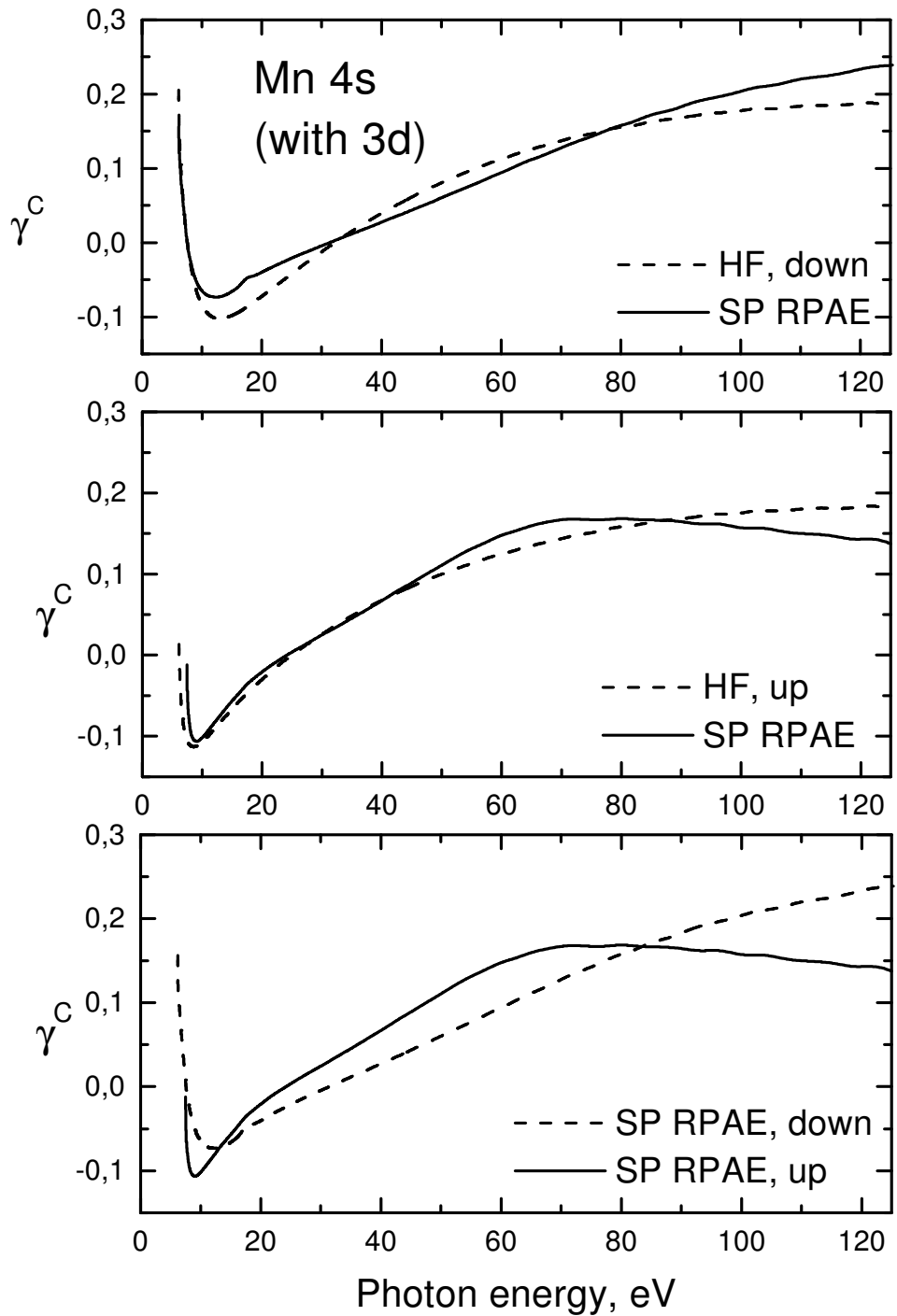

**Figure 7.** Non-dipole parameter $\gamma^c$ for 4s-electrons in Mn.

SP RPAE takes into account the interaction of 4s with 3d-electrons. Solid line is SP RPAE and dashed line is HF results, respectively. $I_{4s\uparrow} = 0.5469$ Ry, $I_{4s\downarrow} = 0.4521$ Ry.

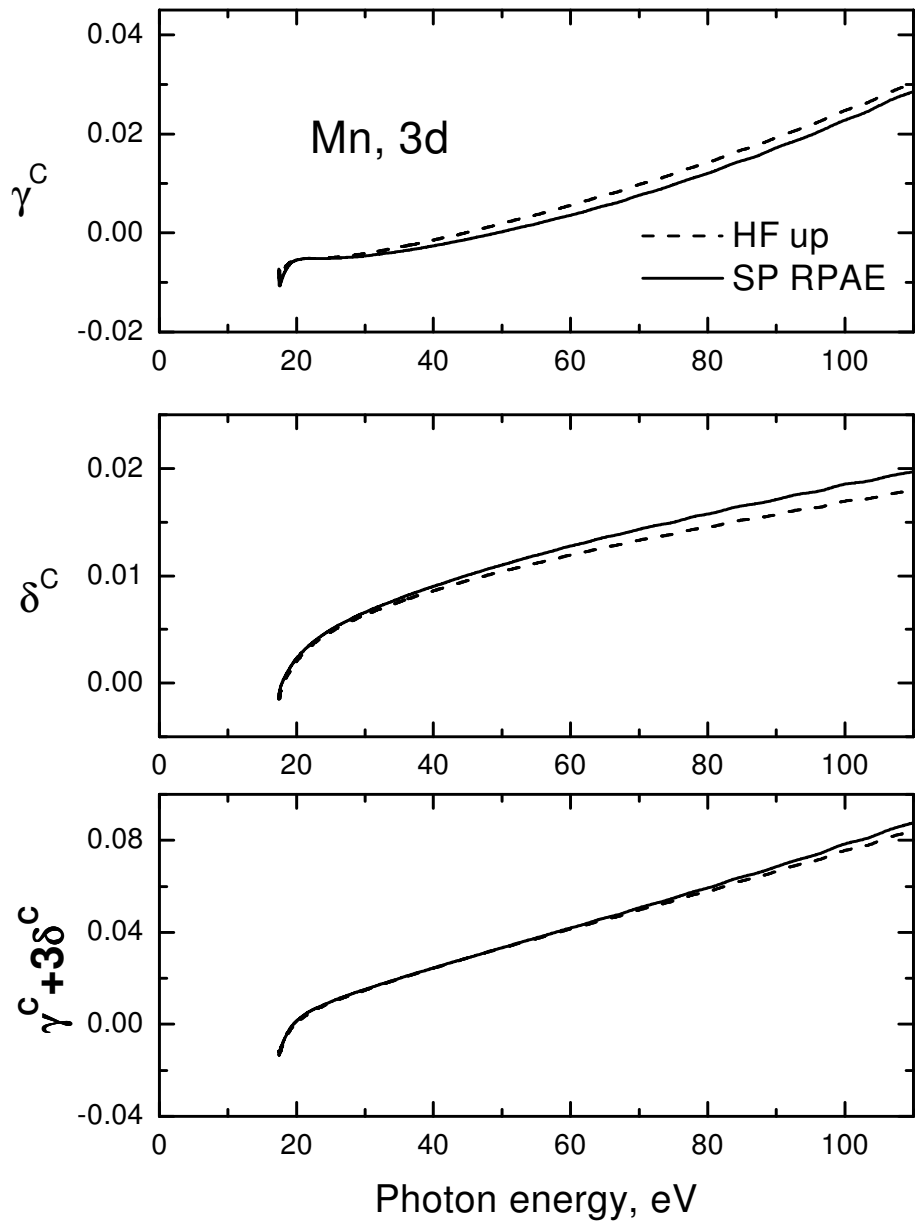

**Figure 8.** Non-dipole parameters for $3d\uparrow$-electrons in Mn. Solid line is SP RPAE and dashed line is HF results, respectively. $I_{3d\uparrow} = 1.2817$ Ry.

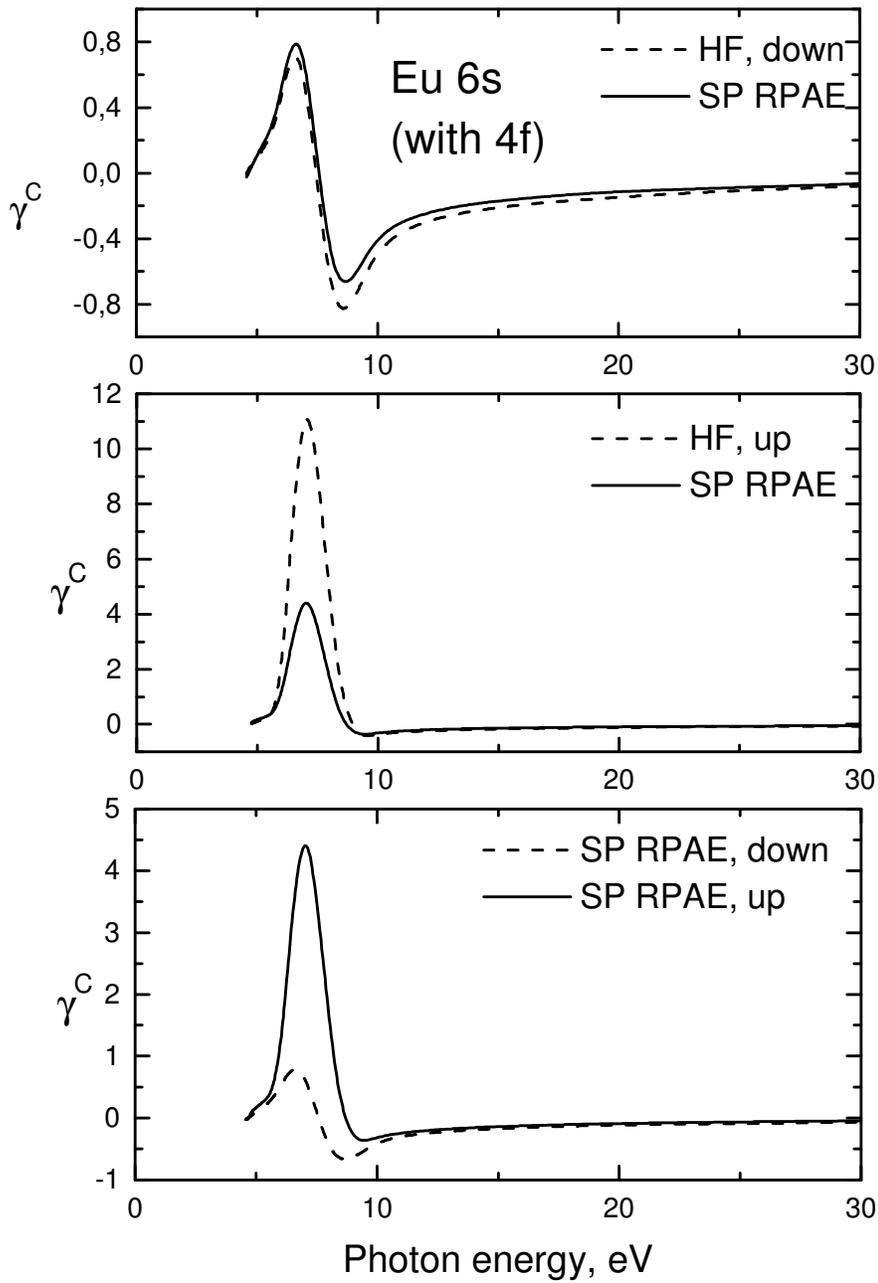

**Figure 9.** Non-dipole parameter $\gamma^c$ for 6s-electrons in Eu.
SP RPAE takes into account the ibnteraction of 6s with 4f-electrons. Solid line is SP RPAE and dashed line is HF results, respectively. $I_{6s\uparrow} = 0.349$ Ry, $I_{6s\downarrow} = 0.335$ Ry, $I_{4f\uparrow} = 1.429$ Ry.

## Acknowledgments

The authors are grateful to the Israeli Science Foundation, Grant 174/03 for financial support.